\documentclass[12pt]{article}
\usepackage{epsf}
\usepackage[dvips]{color}
\usepackage{exscale}
\usepackage[dvips]{graphicx}
\usepackage{amsmath}
\usepackage{amssymb}
\usepackage{latexsym}

\newcommand{\lseq}{\hspace{0.3em}\raisebox{0.4ex}{$<$}\hspace{-0.75em}\raisebox{-.7ex}{$\sim$}\hspace{0.3em}}
\newcommand{\gseq}{\hspace{0.3em}\raisebox{0.4ex}{$>$}\hspace{-0.75em}\raisebox{-.7ex}{$\sim$}\hspace{0.3em}}

\begin{document}
\setlength{\baselineskip}{18pt}
\begin{titlepage}
\begin{flushright}
KUNS-2235\\
YITP-09-62
\end{flushright}

\vspace*{1.2cm}
\begin{center}
{\Large\bf Fine Tuning in General Gauge Mediation}
\end{center}
\lineskip .75em
\vskip 1.5cm

\begin{center}
{\large 
Tatsuo Kobayashi$^{a,}$\footnote[1]{E-mail:
\tt kobayash@gauge.scphys.kyoto-u.ac.jp}, 
Yuichiro Nakai$^{b,}$\footnote[2]{E-mail:
\tt ynakai@yukawa.kyoto-u.ac.jp}, 
and 
Ryo Takahashi$^{b,c,}$\footnote[3]{E-mail: 
\tt Ryo.Takahashi@mpi-hd.mpg.de}
}\\

\vspace{1cm}

$^a${\it Department of Physics, Kyoto University, Kyoto 606-8502, Japan}\\
$^b${\it Yukawa Institute for Theoretical Physics, Kyoto University, Kyoto 
606-8502, Japan}
$^c${\it Max-Planck-Institute f$\ddot{u}$r Kernphysik, Postfach 10 39 80, 69029
Heidelberg, Germany}\\

\vspace*{10mm}
{\bf Abstract}\\[5mm]
{\parbox{13cm}{\hspace{5mm}
%%%%%%%%%%%%%%%%%%%%%%%%%%%%%%%%%%%%%%%%%%%%%%%%%%%%%%%%%%%%%%%%
%             ABSTRACT                                         %
%%%%%%%%%%%%%%%%%%%%%%%%%%%%%%%%%%%%%%%%%%%%%%%%%%%%%%%%%%%%%%%%
We study the fine-tuning problem in the context of general gauge 
mediation. 
Numerical analyses toward for relaxing fine-tuning 
 are presented. We analyse the problem in typical three cases of the messenger 
scale, that is, GUT ($2\times10^{16}$ GeV), intermediate ($10^{10}$ GeV), and 
relatively low energy ($10^6$ GeV) scales. 
In each messenger scale, the parameter space 
reducing the degree of tuning as around $10\%$ is found. 
Certain ratios among gluino mass, wino mass and soft scalar masses are 
favorable.
It is shown that the favorable 
region becomes narrow as the messenger scale becomes lower, and 
tachyonic initial conditions of stop masses at the messenger scale 
are favored to relax the fine-tuning problem  
for the relatively low energy messenger scale.  
Our spectra would also be important from the viewpoint of 
the $\mu-B$ problem.
}}
\end{center}
\end{titlepage}

%%%%%%%%%%%%%%%%%%%%%%
\section{Introduction}
%%%%%%%%%%%%%%%%%%%%%%

Low-energy supersymmetric extension of the standard model is 
one of promising candidates for a new physics at a TeV scale.
The supersymmetry (SUSY) can stabilize the huge hierarchy 
between the weak scale and 
the Planck scale.
That is a motivation for the low-energy SUSY.
In addition, the three gauge couplings are unified 
at the grand unified theory (GUT) scale, 
$2 \times 10^{16}$ GeV, in the minimal supersymmetric 
standard model (MSSM).
Also, supersymmetric standard models have candidates for the 
dark matter.

Although low-energy SUSY solves the (huge) hierarchy problem 
between the weak scale and Planck/GUT scale, 
a few percent of fine-tuning is required in the MSSM as follows.
The lightest CP-even Higgs mass $m_h$ is 
predicted as $m_h \lesssim M_Z$ at the tree level in the MSSM, but that is 
smaller than the experimental bound $m_h \gtrsim 114.4$ GeV.
However, the Higgs mass receives a large radiative correction 
depending on the averaged stop mass 
$m_{\tilde t}$~\cite{Okada:1990gg,Carena:1995wu}.
The experimental bound $m_h \gtrsim 114.4$ GeV requires 
$m_{\tilde t} \gtrsim 1$ TeV when $|A_t|/m_{\tilde t} \lesssim 1.0$, 
where $A_t$ is the so-called A-term corresponding to the top Yukawa 
coupling.
On the other hand, the stop mass also has a renormalization group (RG)
effect on  the soft scalar mass $m_{H_u}$ 
of the up-sector Higgs field as~\cite{Inoue:1982ej,Ibanez:1983di} 
\begin{eqnarray}\label{radiative}
\Delta m_{H_u}^2 \sim -\frac{3y_t^2}{4 \pi^2} m_{\tilde t}^2 \ln
\frac{\Lambda}{m_{\tilde t}},
\end{eqnarray}
where $y_t$ is the top Yukawa coupling and $\Lambda$ denotes 
a cut-off scale of the MSSM such as the Planck scale or GUT scale.
This RG effect $|\Delta m_{H_u}^2|$ would be comparable to 
the stop mass with a negative sign.
Furthermore, the successful electroweak (EW) symmetry breaking 
requires 
\begin{eqnarray}
\frac{1}{2} M_Z^2 \sim - \mu^2 - m_{H_u}^2 ,
\end{eqnarray}
where $\mu$ denotes the supersymmetric mass of the up-sector Higgs 
field $H_u$ and the down-sector Higgs field $H_d$. 
If $m_{H_u}^2 \sim - m_{\tilde t}^2$ and $m_{\tilde t} ={\cal O}(1)$
TeV, one needs a few percent of fine-tuning between $\mu^2$ and 
$m_{H_u}^2$ in order to derive the correct value of $M_Z$.
That is the so-called little hierarchy problem~\cite{Barbieri:1987fn}.
Several works have been done to address this issue~\cite{
Brignole:2003cm}-\cite{Graham}.
%Batra:2003nj,Harnik:2003rs,Kobayashi:2004pu,Kobayashi:2006fh,
%Choi:2005hd,Dermisek:2006ey,Dermisek:2006qj,Gogoladze:2009bd,Horton:2009ed,
%Dermisek:2009si}.
Some of them include extensions of the MSSM.

In the bottom-up approach~\cite{Abe:2007kf}, it is found that 
non-universal gaugino masses with a certain ratio 
are favorable to improve fine-tuning 
in the MSSM when the messenger scale of SUSY breaking 
is the Planck/GUT scale.
Such a favorable ratio of gaugino masses can be realized in 
the TeV scale mirage mediation~\cite{Choi:2004sx,
Choi:2005uz,Choi:2005hd,Choi:2006xb}  and gravity mediation, e.g. 
moduli mediation~\cite{Brignole:1993dj,Horton:2009ed} and 
the SUSY breaking scenario, where 
F-components of gauge non-singlets are sizable~\cite{Ellis:1984bm,
Martin:2009ad,Gogoladze:2009bd} \footnote{
Those spectra with less fine-tuning also have interesting 
aspects on the dark matter physics~\cite{Bae:2007pa}.}.
On the other hand, the spectrum of the constrained MSSM with 
the universal gaugino mass would be unfavorable.
It is also pointed out that a negative value of the stop mass squared 
at the Planck/GUT scale would also be favorable~\cite{
Dermisek:2006ey,Dermisek:2006qj}.

Since the minimal gauge mediation~\cite{Giudice:1998bp} leads to the universal 
gaugino mass, that would be unfavorable from the viewpoint 
of fine-tuning~\cite{Kobayashi:2006fh,Cheung:2007es}.
Recently, Meade, Seiberg and Shih have extended the gauge mediation to 
general gauge mediation (GGM)~\cite{Meade:2008wd}.
(See also \cite{Carpenter:2008wi}-\cite{Buican:2009vv}.)
That leads to non-universal gaugino and soft scalar masses.
Thus, it is important to study fine-tuning in the GGM.
That is our purpose.~\footnote{
See also \cite{Cheung:2007es,Carpenter:2008he}.}
The important difference of the gauge mediation 
(including GGM) from other mediation scenarios 
such as gravity mediation is that the 
messenger scale can vary from the GUT scale to  a TeV scale and 
predicted A-terms are very small in most of models.
These would also lead to an important difference in 
the fine-tuning behavior.

This paper is organized as follows.
In section 2, we briefly review on the fine-tuning problem 
in the MSSM.
Section 3 is also a brief review on the GGM.
In section 4, we analyse numerically on fine-tuning 
in the GGM.
In section 5, we give a comment on the $\mu-B$ problem.
Section 6 is devoted to conclusion.

%%%%%%%%%%%%%%%%%%%%%%%%%%%%%%%%%%%%%%%%%%%%%%%%%%%%%%%%%%%%%
%\section{The fine-tuning probelm in General Gauge Mediation}
%%%%%%%%%%%%%%%%%%%%%%%%%%%%%%%%%%%%%%%%%%%%%%%%%%%%%%%%%%%%%

\section{Fine tuning in the MSSM}

Here, we briefly review the fine-tuning problem in the MSSM by 
showing explicitly equations. 
In our analysis, we neglect the Yukawa couplings except 
the top Yukawa coupling $y_t$.
Then, the Higgs sector in the MSSM is described as the following superpotential,
 \begin{eqnarray}
  W_{\mbox{{\scriptsize Higgs}}}=\mu H_uH_d+y_tQ_3U_3H_u,
 \end{eqnarray}
where $Q_3$, and $U_3$ are the chiral 
superfields corresponding to the left- and right-handed
 top quarks, respectively. 
The Higgs fields and top-stop multiplets as well as 
the gaugino fields play an important role in 
the fine-tuning problem.
Thus, we concentrate on these fields.
Their soft SUSY breaking terms are given by
 \begin{eqnarray}
  V_{\mbox{{\scriptsize soft}}}
   &=& m_{H_u}^2|H_u|^2+m_{H_d}^2|H_d|^2+m_{Q_3}^2|Q_3|^2+m_{U_3}^2|U_3|^2 
       \nonumber \\
   & & +(\mu BH_uH_d+y_tA_tQ_3U_3H_u+\mbox{h.c.}),
 \end{eqnarray}
where $m_X$ $(X=H_{u,d},Q_3,U_3)$ are the soft scalar masses for $X$, 
respectively, $\mu B$ is the SUSY breaking mass, 
i.e. the so-called $B$-term.
Note that we utilize the same notation for denoting a 
chiral superfield and its lowest scalar component.

The soft SUSY 
breaking mass for the up-type Higgs $m_{H_u}$ is subject to relatively large 
logarithmic radiative correction (\ref{radiative}) from mainly stop loops. 
The radiative correction $\Delta m^2_{H_u}$ is comparable to the 
stop mass with the negative sign, i.e. 
$\Delta m^2_{H_u} \sim -m_{\tilde{t}}^2$. 
Such a large and negative correction leads to the EW
symmetry breaking at the weak scale.
Here, we define the averaged top squark mass
$m_{\tilde{t}}$  as
 \begin{eqnarray}
  m_{\tilde{t}}^2\equiv\sqrt{m_{Q_3}^2(M_Z)m_{U_3}^2(M_Z)}. \label{MZ}
 \end{eqnarray}

A stationary condition of the
 Higgs potential gives the relation among the $Z$ boson mass $M_Z$, 
the $\mu$ parameter and soft scalar masses,  
$m_{H_u}^2$ and $m_{H_d}^2$, as
 \begin{eqnarray}
  \frac{M_Z^2}{2}
   =-\mu^2(M_Z)-\frac{m_{H_u}^2(M_Z)\tan^2\beta-m_{H_d}^2(M_Z)}{\tan^2\beta-1},
  \label{Z-boson-mass}
 \end{eqnarray}
where $\tan\beta\equiv\langle H_u\rangle/\langle H_d\rangle$. 
The lightest Higgs boson mass is constrained by
 \begin{eqnarray}
  m_h^2 &\leq& M_Z^2\cos^22\beta\left(1-\frac{3m_t^2}{8\pi^2v^2}\ln
                                      \frac{m_{\tilde{t}}^2}{m_t^2}\right)
               \nonumber \\
        &    & +\frac{3m_t^4}{4\pi^2v^2}
                \bigg[\ln\frac{m_{\tilde{t}}^2}{m^2_t}
                      +\frac{\tilde{A}_t^2}{m_{\tilde{t}}^2}
                       \bigg(1-\frac{\tilde{A}_t^2}{12m_{\tilde{t}}^2}\bigg)
                \nonumber \\
        &    & +\frac{1}{16\pi^2}\bigg(\frac{3m_t^2}{2v^2}
                                       -32\pi\alpha_3\bigg) 
                \bigg\{\frac{2\tilde{A}_t^2}{m_{\tilde{t}}^2}
                       \bigg(1-\frac{\tilde{A}_t^2}{12m_{\tilde{t}}^2}
                       \bigg)\ln\frac{m_{\tilde{t}}^2}{m_t^2}
                       +\bigg(\ln\frac{m_{\tilde{t}}^2}{m_t^2}\bigg)^2
                \bigg\}\bigg], \label{lightest-Higgs-mass}
 \end{eqnarray}
within the 2-loop approximation \cite{Carena:1995wu}, where 
$v=174$ GeV, 
$\tilde{A}_t\equiv A_t(M_Z)-\mu\cot\beta$ and 
$m_t$ is the running top squark mass at $M_Z$.

The current experimental lower bound for the Higgs mass is given by the LEP 
experiment as $m_h\geq114.4$ GeV. 
%The 2-loop approximation for the Higgs mass 
%\eqref{lightest-Higgs-mass} suggest that the ratio of the $A$-term to the top 
%squark mass must be larger than of order one, 
%$|A_t(M_Z)/m_{\tilde{t}}|\gseq\mathcal{O}(1)$ \cite{Abe:2007kf}. 
In order to realize $m_h\geq114.4$ GeV, 
a large top squark mass is required as 
$m_{\tilde{t}}\gseq 1$ TeV when $|A_t(M_Z)/m_{\tilde{t}}|\lseq 1.0$.
The soft scalar mass of the up-sector Higgs field, $m_{H_u}$ 
suffers from a large 
radiative correction according to such a large top squark mass through 
\eqref{radiative}.
Therefore, a few percent of fine-tuning between 
$m_{H_u}^2$ and $\mu^2$ is 
required in \eqref{Z-boson-mass} 
in order to realize the EW symmetry breaking with the 
experimentally observed $Z$ boson mass, $M_Z\simeq91.2$ GeV. 
That is  the so-called little hierarchy problem.
We investigate this fine-tuning problem in the context of the GGM.
Furthermore, when $|A_t(M_Z)/m_{\tilde{t}}|\lseq1.5$, 
the condition $m_h\geq114.4$ GeV requires $m_{\tilde{t}}\gseq500$ GeV.
Hence, the stop mixing $A_t/m_{\tilde{t}}$ 
is important \cite{Carena:1995wu,Essig:2007vq}.

\section{General gauge mediation}

Before considering the fine-tuning problem in the GGM, we also give a 
brief review on the GGM. 
Recently, Meade, Seiberg and Shih have presented the most 
general spectrum which can be obtained in gauge mediated SUSY breaking model 
\cite{Meade:2008wd}. A careful definition of gauge mediation mechanism has been
 given in the work, that is, in the limit that the MSSM gauge couplings 
$\alpha_i\rightarrow0$, the theory decouples into the MSSM and a separate 
hidden sector which breaks SUSY. 
Following the convention, we label the gauge groups, 
$SU(3)$, $SU(2)$ and $U(1)$ of the MSSM by $a=3,2,1$, respectively.
Within the framework of the GGM, the three gaugino masses $M_a$ 
($a=1,2,3$) of the MSSM are given 
at the messenger scale $M$ as,
 \begin{eqnarray}
  M_a             &=& 2 g_a^2 {B}_{a}.
\end{eqnarray}
In general, ${B}_{a}$ ($a=1,2,3$) are three independent complex parameters.
If CP phases of ${B}_{a}$ are not aligned each other, 
that would lead to a serious CP problem.
Thus, we use ${B}_{a}$ as three real parameters.
The soft scalar masses squared are also given in the GGM as 
\begin{eqnarray}
  m_{\tilde{f}}^2 &=& g_1^2Y_f\zeta+\sum_{a=1}^3g_a^4c_2(f;a)A_a,
 \end{eqnarray}
at $M$,
where $c_2(f;a)$ is the quadratic Casimir of the representation of fermion $f$ 
under the gauge group corresponding to the label $a$.
Here, $A_a$ ($a=1,2,3$) are three independent real parameters.
Hereafter, we concentrate on the models with $\zeta=0$.\footnote{
This situation, $\zeta =0$, can be realized by invoking messenger parity.}
In this case, there are the mass relations at $M$
 \begin{eqnarray}
  m_{Q_f}^2+m_{D_f}^2+m_{E_f}^2-m_{L_f}^2-2m_{U_f}^2=0, 
\qquad m_{H_u}=m_{H_d},
\end{eqnarray}
 where $m_{Q_f}$, $m_{U_f}$, $ m_{D_f}$, $m_{L_f}$, and $m_{E_f}$
 denote soft scalar masses for  
the $f$-th generation of the left-handed squarks, up-sector right-handed 
squarks, down-sector right-handed squarks, left-handed sleptons and 
right-handed sleptons.
Thus, the $U(1)_Y$ $D$-term $S$, i.e., 
\begin{eqnarray}
  S=m_{H_u}^2-m_{H_d}^2+\sum_{f=1}^3(m_{Q_f}^2+m_{D_f}^2+m_{E_f}^2-m_{L_f}^2
    -2m_{U_f}^2), \label{S}
 \end{eqnarray}
vanishes at the messenger scale $M$.
Furthermore, its RG equation is given as 
\begin{eqnarray}
  (4 \pi)^2\frac{dS}{dt}=-b_1g^2_1(t)S(t),  
 \end{eqnarray}
where $t\equiv2\log(M_Z/\bar{\mu})$, $\bar{\mu}$ is an arbitrary energy scale, 
and $b_1=33/5$ (and $b_2=1$, $b_3=-3$ for references). 
Thus, when $S$ is vanishing at $M$, it vanishes 
at any scale.
For concreteness, we show explicitly the initial conditions of 
the soft scalar masses, 
$m_{Q_3}$, $m_{U_3}$, $m_{H_u}$ and $m_{H_d}$ as 
 \begin{eqnarray}
  m_{Q_3}^2(M) 
   &=& \frac{4}{3}g_3^4(M)A_3+\frac{3}{4}g_2^4(M)A_2
       +\frac{3}{5}\bigg(\frac{1}{6}\bigg)^2g_1^4(M)A_1 \nonumber \\
   &=& (4\pi)^4B_3^2
       \bigg[\frac{4}{3}\tilde{\alpha}_3^2(M)a_3
             +\frac{3}{4}\tilde{\alpha}_2^2(M)a_2
             +\frac{1}{60}\tilde{\alpha}_1^2(M)a_1
       \bigg], \nonumber\\
  m_{U_3}^2(M) 
   &=& \frac{4}{3}g_3^4(M)A_3+\frac{3}{5}\bigg(-\frac{2}{3}\bigg)^2g_1^4(M)A_1 
       \nonumber\\
   &=& (4\pi)^4B_3^2
       \bigg[\frac{4}{3}\tilde{\alpha}_3^2(M)a_3
             +\frac{4}{15}\tilde{\alpha}_1^2(M)a_1\bigg], \\
  m_{H_u}^2(M)
   &=& m_{H_d}^2(M) \nonumber\\
   &=& \frac{3}{4}g_2^4(M)A_2
       +\frac{3}{5}\bigg(\pm\frac{1}{2}\bigg)^2g_1^4(M)A_1 \nonumber\\
   &=& (4\pi)^4B_3^2
       \bigg[\frac{3}{4}\tilde{\alpha}_2^2(M)a_2
             +\frac{3}{20}\tilde{\alpha}_1^2(M)a_1
       \bigg], \nonumber
 \end{eqnarray}
where $\tilde{\alpha}_a\equiv\alpha_a/(4\pi)\equiv g_a^2/(4\pi)^2$.
Here, we have defined the ratios
\begin{eqnarray}
a_a \equiv \frac{A_a}{B_3^2},
\end{eqnarray}
for convenience.
Similarly, we define the ratios of gaugino masses to the 
gluino mass, 
\begin{eqnarray}
b_a \equiv \frac{B_a}{B_3}.
\end{eqnarray}

The initial condition of the $A$-term in the GGM is given as 
\begin{eqnarray}
A_t =0,
\end{eqnarray}
at $M$.
Thus, the $A$-term $A_t$ at the weak scale is given only 
by the RG effect between the weak scale and the messenger scale $M$.
This initial condition is important because 
the stop mixing $A_t/m_{\tilde t}$ at the weak scale 
has a significant effect on the Higgs mass (\ref{lightest-Higgs-mass}).

By utilizing these gaugino and sfermion masses given in the GGM, we numerically
 analyze the fine-tuning problem in the next section. 

%%%%%%%%%%%%%%%%%%%%%%%%%%%%
\section{Numerical Analyses}
%%%%%%%%%%%%%%%%%%%%%%%%%%%%

We study the fine-tuning problem in the GGM and present numerical 
analyses. 
In gauge mediated SUSY breaking models, phenomenological consequences
 at the EW scale generally depend on the messenger scale $M$. 
We present our analyses for three typical messenger scales, 
that is (i) GUT scale 
$M=\Lambda_{\mbox{{\scriptsize GUT}}}\equiv2\times10^{16}$ GeV, (ii) 
intermediate scale $M=10^{10}$ GeV, and (iii) relatively low energy scale 
$M=10^6$ GeV.

Firstly, we give the soft parameters at the EW scale by integrating the 1-loop 
RG equations \cite{Inoue:1982ej}. The gaugino mass at 
the EW scale are
 \begin{eqnarray}
  M_1(M_Z) &\simeq& 0.428B_1, \\
  M_2(M_Z) &\simeq& 0.859B_2, \\
  M_3(M_Z) &\simeq& 3.00B_3. 
 \end{eqnarray} 
In this analysis, we use the values of gauge couplings at the EW scale as
$\tilde{\alpha}_1(M_Z)\simeq1.36\times10^{-3}$, 
$\tilde{\alpha}_2(M_Z)\simeq2.72\times10^{-3}$, and 
$\tilde{\alpha}_3(M_Z)\simeq9.50\times10^{-3}$.
These couplings in the MSSM 
would be unified at the GUT scale within a good accuracy. 
In addition, we use the running top mass $m_t=164.5$ GeV at $M_Z$ 
and $\tan \beta = 10$ for numerical analysis.

The scalar masses such as 
$m_{Q_3}$, $m_{U_3}$, $m_{H_{u,d}}$, and $A_t$, which are important to discuss 
the fine-tuning problem, are given for each typical messenger scale as

\noindent(i)  $M=\Lambda_{\mbox{{\scriptsize GUT}}}$,
 \begin{eqnarray}
  m_{Q_3}^2(M_Z) &\simeq& 6.07B_3^2-0.0120B_1B_3-0.00754B_1^2
                          -0.0834B_2B_3 \nonumber \\
                 &      & -0.00245B_1B_2+0.437B_2^2 \nonumber \\
                 &      & -0.116m_{H_u}^2(M)+0.884m_{Q_3}^2(M)
                          -0.116m_{U_3}^2(M), \label{GUT-Q3}\\
  m_{U_3}^2(M_Z) &\simeq& 5.11B_3^2-0.0240B_1B_3+0.0495B_1^2
                          -0.167B_2B_3 \nonumber \\
                 &      & -0.00490B_1B_2-0.202B_2^2 \nonumber \\
                 &      & -0.232m_{H_u}^2(M)-0.232m_{Q_3}^2(M)
                          +0.768m_{U_3}^2(M), \\
  m_{H_u}^2(M_Z) &\simeq& -2.90B_3^2-0.0361B_1B_3+0.00505B_1^2
                          -0.250B_2B_3 \nonumber \\
                 &      & -0.00735B_1B_2+0.235B_2^2
                          \nonumber \\
                 &      & +0.652m_{H_u}^2(M)-0.348m_{Q_3}^2(M)
                          -0.348m_{U_3}^2(M), \label{GUT-up-Higgs} \\
  m_{H_d}^2(M_Z)  &\simeq& 0.538B_2^2+0.0415B_1^2+m_{H_d}^2(M), 
                          \label{GUT-mhd}\\
  A_t(M_Z)       &\simeq& 2.20B_3+0.278B_2+0.0352B_1,\label{GUT-A}
 \end{eqnarray}
(ii)  $M=10^{10}$ GeV,
 \begin{eqnarray}
  m_{Q_3}^2(M_Z) &\simeq& 5.43B_3^2-0.00327B_1B_3-0.000940B_1^2
                          -0.0331B_2B_3 \nonumber \\
                 &      & -0.000404B_1B_2+0.227B_2^2 \nonumber \\ 
                 &      & -0.0958m_{H_u}^2(M)+0.904m_{Q_3}^2(M)
                          -0.0958m_{U_3}^2(M), \\
  m_{U_3}^2(M_Z) &\simeq& 4.76B_3^2-0.00654B_1B_3+0.0142B_1^2
                          -0.0661B_2B_3 \nonumber \\
                 &      & -0.000807B_1B_2-0.0701B_2^2
                          \nonumber \\
                 &      & -0.192m_{H_u}^2(M)-0.192m_{Q_3}^2(M)
                          +0.808m_{U_3}^2(M), \\
  m_{H_u}^2(M_Z) &\simeq& -2.03B_3^2-0.00981B_1B_3+0.00405B_1^2
                          -0.0992B_2B_3 \nonumber \\
                 &      & -0.00121B_1B_2+0.157B_2^2
                          \nonumber \\
                 &      & +0.712m_{H_u}^2(M)-0.288m_{Q_3}^2(M)
                          -0.288m_{U_3}^2(M), \label{Inter-up-Higgs} \\
  m_{H_d}^2(M_Z)  &\simeq& 0.262B_2^2+0.0103B_1^2+m_{H_d}^2(M), 
                         \label{10-mhd}\\
  A_t(M_Z)       &\simeq& 1.93B_3+0.181B_2+0.0167B_1,\label{10-A}
 \end{eqnarray}
(iii)  $M=10^6$ GeV,
 \begin{eqnarray}
  m_{Q_3}^2(M_Z) &\simeq& 4.24B_3^2-0.000733B_1B_3-0.0000353B_1^2
                          -0.00879B_2B_3 \nonumber \\
                 &      & -0.0000603B_1B_2+0.111B_2^2
                          \nonumber \\
                 &      & -0.0669m_{H_u}^2(M)+0.933m_{Q_3}^2(M)
                          -0.0669m_{U_3}^2(M), \\
  m_{U_3}^2(M_Z) &\simeq& 3.90B_3^2-0.00147B_1B_3+0.00563B_1^2
                          -0.0176B_2B_3 \nonumber \\
                 &      & -0.000121B_1B_2-0.0198B_2^2 
                          \nonumber \\
                 &      & -0.134m_{H_u}^2(M)-0.134m_{Q_3}^2(M)
                          +0.866m_{U_3}^2(M), \\
  m_{H_u}^2(M_Z) &\simeq& -1.03B_3^2-0.00220B_1B_3+0.00234B_1^2
                          -0.0264B_2B_3 \nonumber \\
                 &      & -0.000181B_1B_2+0.0914B_2^2
                          \nonumber \\
                 &      & +0.799m_{H_u}^2(M)-0.201m_{Q_3}^2(M)
                          -0.201m_{U_3}^2(M), \label{Rel-Low-up-Higgs}\\
  m_{H_d}^2(M_Z) &\simeq& 0.121B_2^2+0.00366B_1^2+m_{H_d}^2(M), 
                          \label{6-mhd}\\
  A_t(M_Z)       &\simeq& 1.47B_3+0.105B_2+0.00850B_1.   \label{6-A}
 \end{eqnarray}
Here, we have used the initial conditions, $A_t(M)=S(M)=0$.
The change of RG effects between the cases (ii) and (iii) is 
rather drastic compared with one between (i) and (ii).

If all soft parameters are taken as the same order, 
$B_a\sim m_X(M)$, the averaged top squark mass is approximated 
for each messenger scale as
  \begin{eqnarray}
   m_{\tilde{t}}^2\sim
    \left\{
     \begin{array}{lc}
      6.0B_3^2\hspace{5mm}\mbox{ in the case (i)} \\
      5.7B_3^2\hspace{5mm}\mbox{ in the case (ii)} \\
      4.8B_3^2\hspace{5mm}\mbox{ in the case (iii)}
     \end{array}
    \right..
  \end{eqnarray}
For a fixed value of $|A_t(M_Z)/m_{\tilde t}|$, 
a large value of $m_{\tilde{t}}^2$ would be favorable to realize 
the Higgs mass $m_h\geq114.4$ GeV.
That implies that a higher messenger scale would be favorable   
for a fixed value of the gluino mass, i.e. $B_3$.
In order to satisfy the experimental bound for the Higgs mass, the lower bound
 for $B_3$ is roughly estimated as
 \begin{eqnarray}
  B_3\gseq
   \left\{
    \begin{array}{lc}
     200~(410)~\mbox{GeV for }|A_t(M_Z)/m_{\tilde t}|\lseq1.5~(1.0)\mbox{ in the
case (i)} 
     \\
     210~(420)~\mbox{GeV for }|A_t(M_Z)/m_{\tilde t}|\lseq1.5~(1.0)\mbox{ in the
case (ii)} 
     \\
     230~(460)~\mbox{GeV for }|A_t(M_Z)/m_{\tilde t}|\lseq1.5~(1.0)\mbox{ in the
case (iii)}
     \\
    \end{array}
   \right.. \label{Higgs-mass-bound}
 \end{eqnarray}

Furthermore, we can estimate the stop mixing 
$|A_t(M_Z)/m_{\tilde  t}|$.
For example, for $B_a\sim m_X(M)$ we estimate 
 \begin{eqnarray}
  |A_t(M_Z)/m_{\tilde t}| \sim
    \left\{
     \begin{array}{lc}
      1.0\hspace{7mm}\mbox{ in the case (i)} \\
      0.89\hspace{5mm}\mbox{ in the case (ii)} \\
      0.72\hspace{5mm}\mbox{ in the case (iii)}
     \end{array}
    \right..
  \end{eqnarray}
A large value of $|A_t(M_Z)/m_{\tilde t}| $ would be favorable to realize 
the Higgs mass $m_h\geq114.4$ GeV.
That implies that a higher messenger scale would be favorable.

On the other hand, the dominant part of the RG effects 
in $m_{H_d}^2$ \eqref{GUT-up-Higgs}, 
\eqref{Inter-up-Higgs} and \eqref{Rel-Low-up-Higgs} is due to 
the gluino mass, i.e. $B_3^2$.
If $B_3 \sim 500$ GeV, we need fine-tuning between 
$m_{H_u}^2$ and $\mu^2$ to realize $M_Z$.
The absolute value of coefficient of $B_3^2$ in 
 $m_{H_d}^2(M_Z)$ decreases as the messenger scale $M$ 
decreases.
Thus, for a fixed value of $B_3$, 
the degree of fine-tuning is reduced as 
the messenger scale becomes lower.

Thus, the tension between the fine-tuning and the lower bound of 
the Higgs mass $m_h\geq114.4$ GeV depends non-trivially on 
the messenger scale $M$.
Also that would depend on ratios among gaugino masses and 
scalar masses, although we have used $B_a\sim m_X(M)$ 
in the above estimation.

Toward the numerical analyses of the fine-tuning problem, we introduce 
fine-tuning parameters \cite{Barbieri:1987fn},
 \begin{eqnarray}
  \Delta_Y\equiv\frac{1}{2}\frac{Y}{M_Z^2}\frac{\partial M_Z^2}{\partial Y},
 \end{eqnarray}
which indicates that we need $100/\Delta_Y$ percent of fine-tuning 
for $Y$ to derive $M_Z$.
A larger value of $\Delta_Y$ means more severe fine-tuning to be
required.

If $B_a$ and $A_a$ are independent of each other, 
fine-tuning for $B_3$ would be most severe, 
because $m_{H_d}^2(M_Z)$  depends dominantly on $B_3$.
For example, we can calculate

\noindent (i) $M=\Lambda_{\mbox{{\scriptsize GUT}}}$
 \begin{eqnarray}
  \Delta_{B_3} &=& 5.85\hat{M}_3^2+(0.0364\hat{M}_1+0.253\hat{M}_2
                  )\hat{M}_3, 
 \end{eqnarray}
(ii) $M=10^{10}$ GeV
 \begin{eqnarray}
  \Delta_{B_3} &=& 4.10\hat{M}_3^2+(0.00990\hat{M}_1+0.100\hat{M}_2
                 )\hat{M}_3, 
 \end{eqnarray}
(iii) $M=10^6$ GeV
 \begin{eqnarray}
  \Delta_{B_3} &=& 2.08\hat{M}_3^2+(0.00222\hat{M}_1+0.0266\hat{M}_2
                  )\hat{M}_3, 
 \end{eqnarray}
where $\hat{M}_a\equiv B_a/M_Z$. 
It is found  
that the coefficients of the terms become small as the messenger scale 
becomes lower. 
If $\Delta_{B_3}\leq10$ is required under the condition $B_1=B_2=B_3$, the 
allowed value of $B_3$ are (i) $B_3\leq110$ GeV, (ii) $B_3\leq140$ GeV, and 
(iii) $B_3\leq190$ GeV. 
They could not satisfy the bounds on the Higgs mass 
\eqref{Higgs-mass-bound}. 
On the other hand, when we take $B_3 \simeq 500$ 
GeV, we find that severe fine-tunings such as (i) $\Delta_{B_3}\simeq200$, 
(ii) $\Delta_{B_3}\simeq140$, (iii) $\Delta_{B_3}\simeq70$ are needed.

We have assumed that $B_a$ and $A_a$ are independent of each other.
However, in a definite theory, they are not independent, 
but certain ratios are predicted in each theory.
That is, in a definite theory there is one parameter, 
which determines the overall size of soft SUSY breaking terms.
We choose $B_3$ as such a parameter and the ratios 
$a_a$ and $b_a$ are fixed in a theory.
Then, we consider the fine-tuning only for $B_3$, i.e. 
$\Delta_{B_3}$ under fixed ratios of $a_a$ and $b_a$.
Varying $a_a$ and $b_a$ means that we compare different theories 
in the theory space of the GGM.
Then, the fine-tuning parameter can be rewritten as

\noindent (i) $M=\Lambda_{\mbox{{\scriptsize GUT}}}$
 \begin{eqnarray}
    \Delta_{B_3} 
     &=& \hat{M}_3^2(5.85+0.506b_2-0.465b_2^2+0.508a_3-0.122a_2
                              +0.0728b_1+0.0148b_1b_2 \nonumber \\
     & & \phantom{\hat{M}_3^2(}-0.00936b_1^2+0.00132a_1), \label{fine-GUT}
 \end{eqnarray}
(ii) $M=10^{10}$ GeV
 \begin{eqnarray}
  \Delta_{B_3} 
   &=& \hat{M}_3^2(4.10+0.200b_2-0.311b_2^2+0.825a_3-0.143a_2+0.0198b_1
                   +0.00245b_1b_2  \nonumber \\
   & & \phantom{\hat{M}_3^2(}-0.00798b_1^2-0.00495a_1),  
 \end{eqnarray}
(iii) $M=10^6$ GeV
 \begin{eqnarray}
  \Delta_{B_3} &=& \hat{M}_3^2(2.08+0.0533b_2-0.182b_2^2+1.04a_3-0.183a_2
                                +0.00444b_1+0.000365b_1b_2\nonumber \\
                       & & \phantom{\hat{M}_3^2(}-0.00465b_1^2-0.00821a_1).
 \end{eqnarray}
Coefficients of $b_1$ and $a_1$ in the above equations are very small.
Thus, those terms would not be important unless $b_1={\cal O}(10)$
or $a_1={\cal O}(100)$.
Therefore, we concentrate on others and throughout our numerical analyses 
we take $b_1=a_1=1$ as a typical value.
It is found that the coefficients of $a_2$ and $b^2_2$, 
which determines the wino mass, are negative.
Hence, it would be favorable to cancel the dominant term by relatively 
large $b_2$ and/or $a_2$.
%We find that the terms being proportional to $b_2$, $b_2^2$, $a_3$, and $a_2$ 
%appear in the fine-tuning parameter with non-negligible size of coefficient at
% each scale, and the coefficient of $b_2^2$, which determines the wino mass,  
%and $a_2$ are negative. Therefore, there are mainly two method in order to 
%reduce the fine-tuning, that is, canceling the dominant term by relatively 
%large $b_2$ and/or $a_2$. We analyse these two kind of case to relax 
%the little hierarchy problem. 
That is, models satisfying 

\noindent (i) $M=\Lambda_{\mbox{{\scriptsize GUT}}}$
 \begin{eqnarray}
     && |5.85+0.506b_2-0.465b_2^2+0.508a_3-0.122a_2|  \ll 1,
 \end{eqnarray}
(ii) $M=10^{10}$ GeV
 \begin{eqnarray}
& &  |4.10+0.200b_2-0.311b_2^2+0.825a_3-0.143a_2| \ll 1,
 \end{eqnarray}
(iii) $M=10^6$ GeV
 \begin{eqnarray}
 & & |2.08+0.0533b_2-0.182b_2^2+1.04a_3-0.183a_2 | \ll 1, 
 \end{eqnarray}
would be interesting in the theory space.
For fixed values of $a_2$ and $a_3$, a favorable value of 
$b_2$ is determined.
That means a favorable ratio between the gluino and 
wino masses such as Ref.~\cite{Abe:2007kf}.
For a fixed value of $b_2$, a linear correlation between 
$a_2$ and $a_3$ is required.
On the other hand, for a fixed value of $a_2$ ($a_3$) 
a quadratic relation between $b_2$ and $a_3$ ($a_2$) is 
required.

\begin{figure}
\begin{center}
\includegraphics[scale = 0.8]{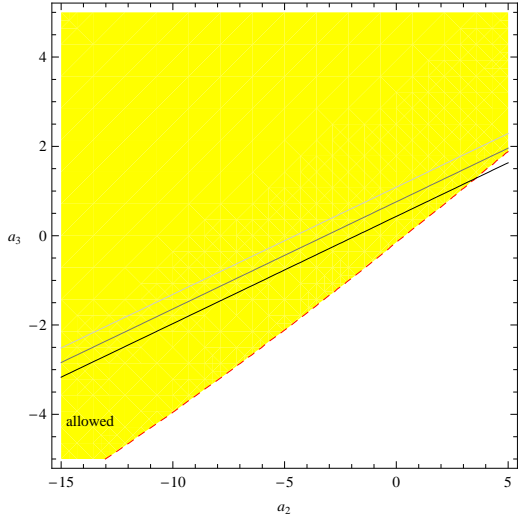}\hspace{1cm}
\includegraphics[scale = 0.8]{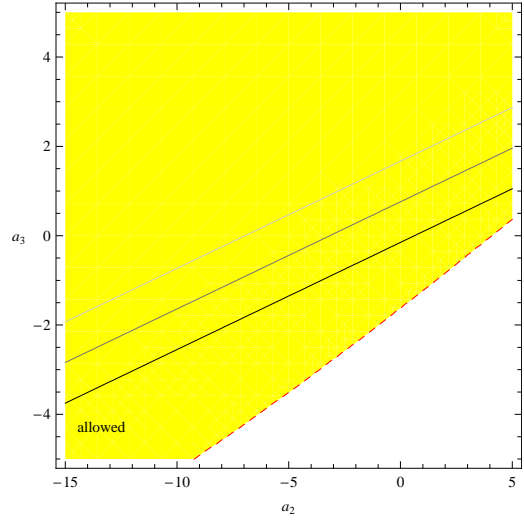}
\end{center}
\hspace{4.3cm}(a)\hspace{7.4cm}(b)

\begin{center}
\includegraphics[scale = 0.8]{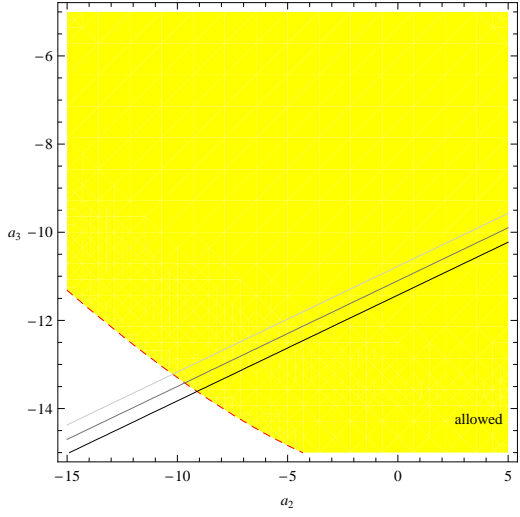}\hspace{1cm}
\includegraphics[scale = 0.8]{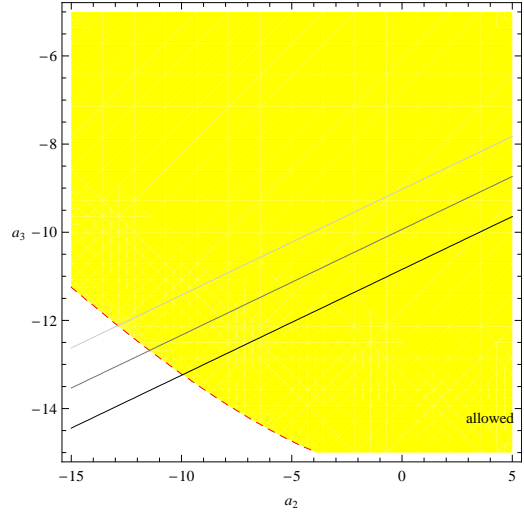}
\end{center} 
\hspace{4.4cm}(c)\hspace{7.4cm}(d)
\caption{Lines and curve for the case (i) determined by constraints from 
$\Delta_{B_3}=5,10,15$ (solid lines), $m_h=114.4$ GeV (dashed (red) curve), and 
experimentally allowed region $m_h\geq114.4$ GeV (shaded (yellow)
region). 
The darker and darkest solid lines correspond $\Delta_{B_3}=10$ and 
$5$, respectively.
We 
take as $b_1=a_1=1$ in all figures. (a) for $B_3=500$ GeV and $b_2\simeq4.19$. 
(b) for $B_3=300$ GeV and $b_2\simeq4.01$. These values of $b_2$ lead to 
$\Delta_{B_3}=10$ when $b_1=a_1=a_2=a_3=1$ in each value of $B_3$. (c) for 
$B_3=500$ GeV and $b_2=1$. (b) for $B_3=300$ GeV and $b_2=1$.} 
\label{fig-2}
\end{figure}

\begin{figure}
\begin{center}
\includegraphics[scale = 0.8]{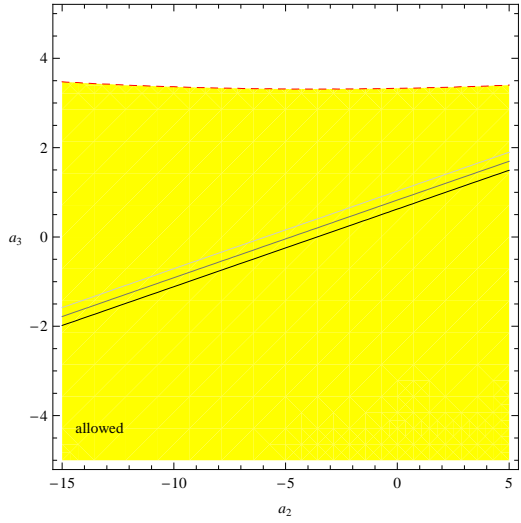}\hspace{1cm}
\includegraphics[scale = 0.8]{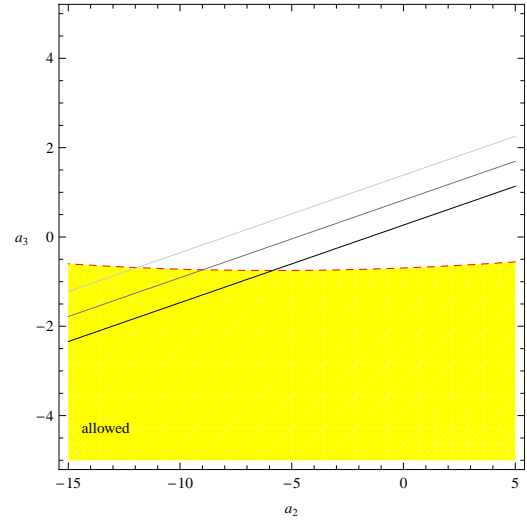}
\end{center}
\hspace{4.3cm}(a)\hspace{7.4cm}(b)

\begin{center}
\includegraphics[scale = 0.8]{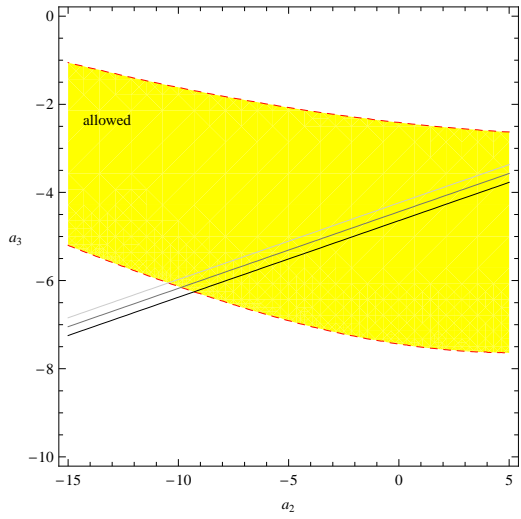}\hspace{1cm}
\includegraphics[scale = 0.8]{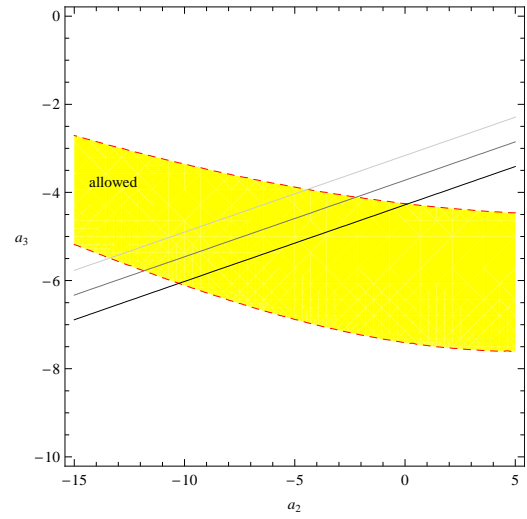}
\end{center}
\hspace{4.4cm}(c)\hspace{7.4cm}(d)
\caption{The same lines and curve as Fig.\ref{fig-2} but in the case (ii)} 
\label{fig-3} 
\end{figure}

\begin{figure}
\begin{center}
\includegraphics[scale = 0.8]{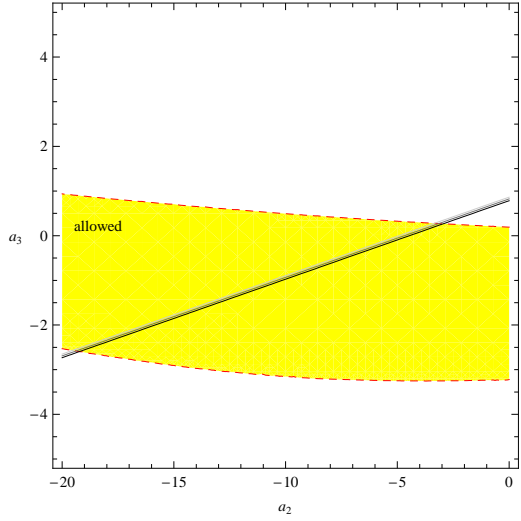}\hspace{1cm}
\includegraphics[scale = 0.8]{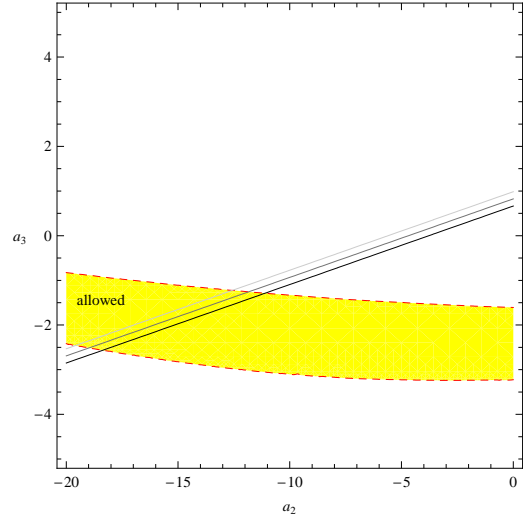}
\end{center}
\hspace{4.3cm}(a)\hspace{7.4cm}(b)

\begin{center}
\includegraphics[scale = 0.8]{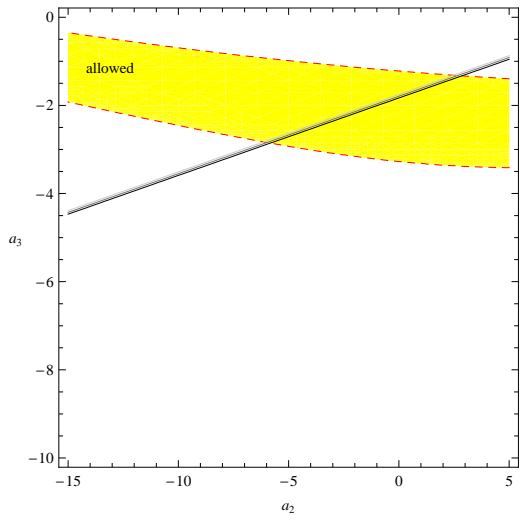}\hspace{1cm}
\includegraphics[scale = 0.8]{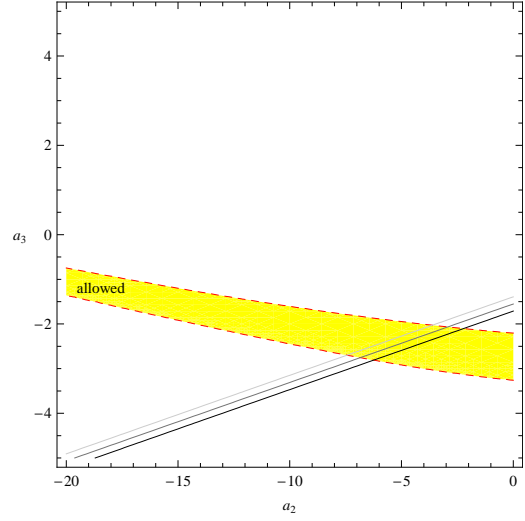}
\end{center}
\hspace{4.4cm}(c)\hspace{7.4cm}(d)
\caption{The same lines and curve as Fig.\ref{fig-2} but (a) and (c) for 
$B_3=1$ TeV and (b) and (d) for $B_3=500$ GeV in the case (iii).}
\label{fig-4}  
\end{figure}

\begin{figure}
\begin{center}
\includegraphics[scale = 0.8]{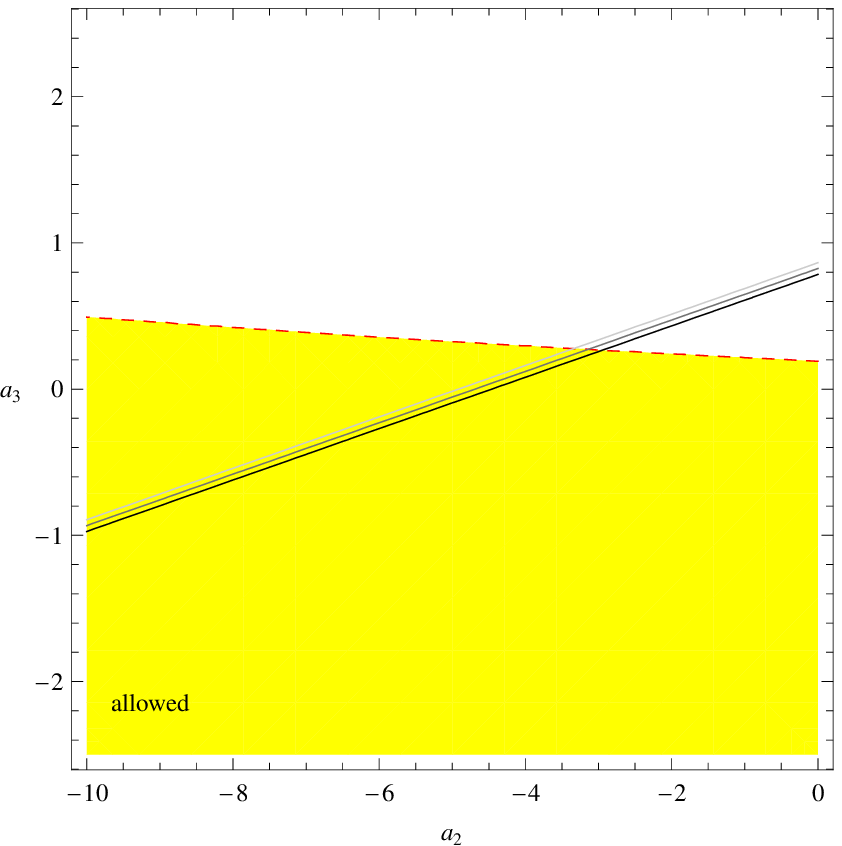}\hspace{1cm}
\includegraphics[scale = 0.8]{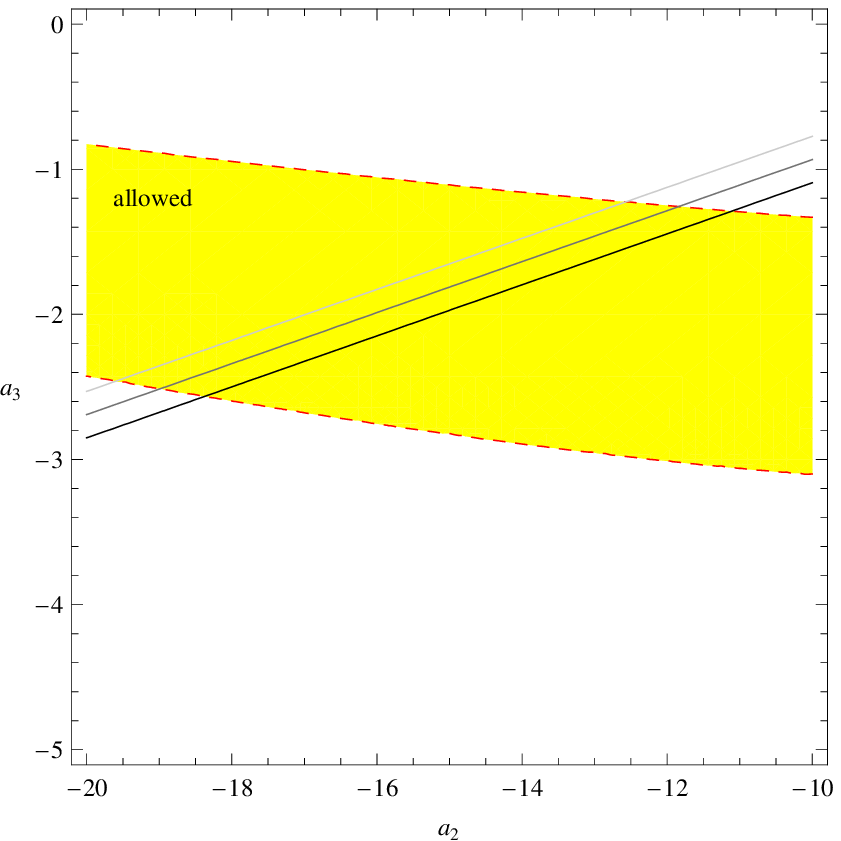}
\end{center}
\hspace{4.3cm}(a)\hspace{7.4cm}(b)

\begin{center}
\includegraphics[scale = 0.8]{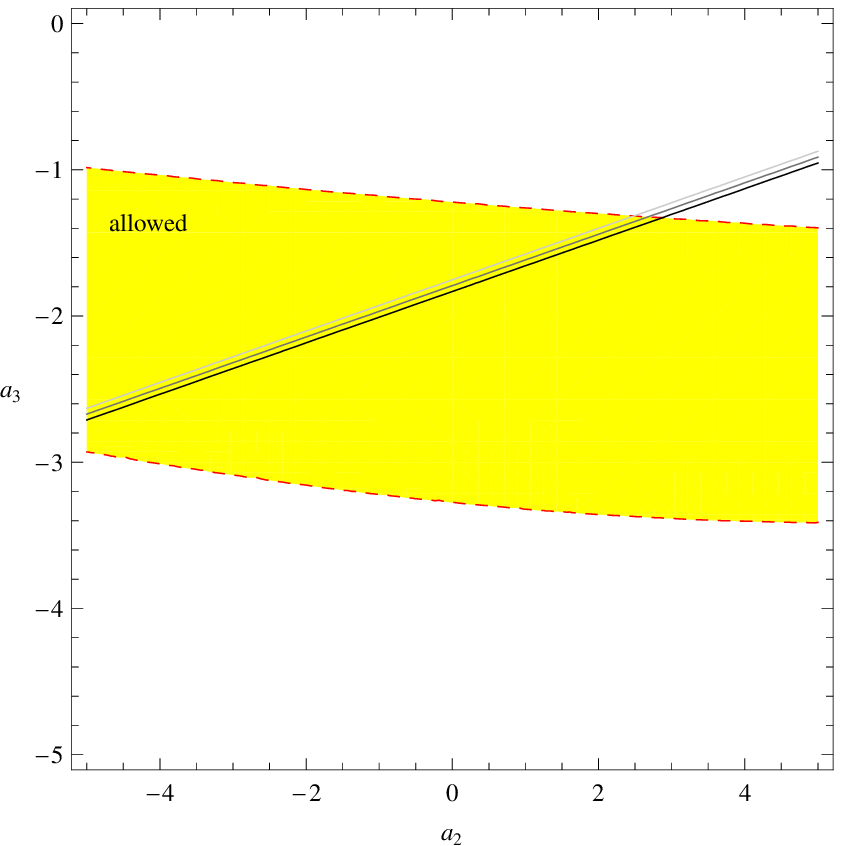}\hspace{1cm}
\includegraphics[scale = 0.8]{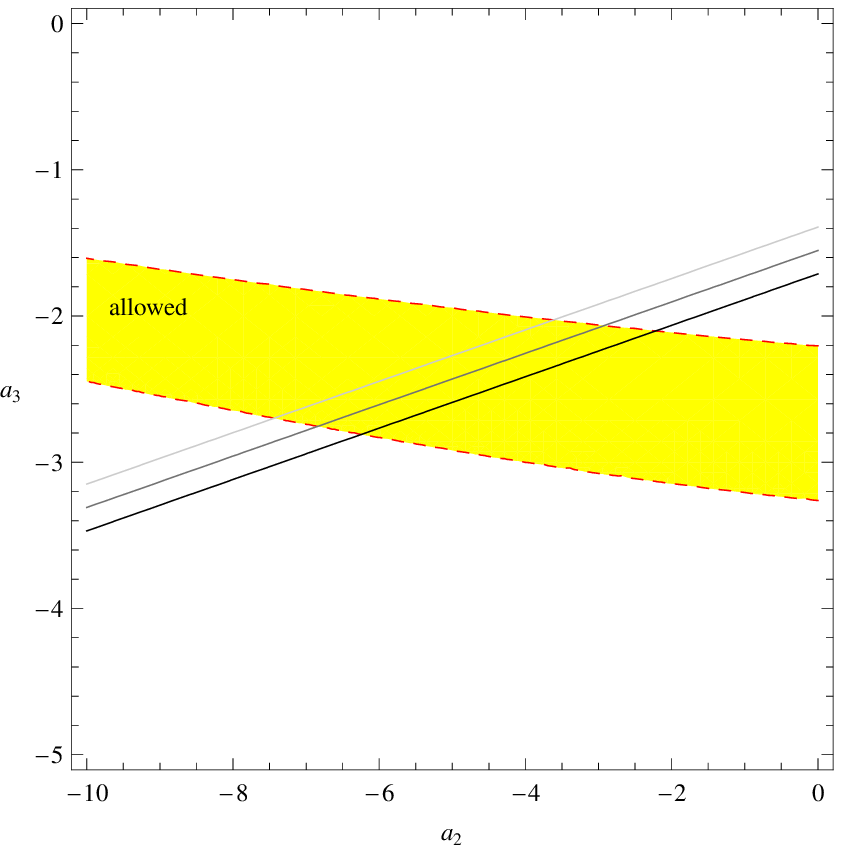}
\end{center}
\hspace{4.4cm}(c)\hspace{7.4cm}(d)
\caption{The enlargements of Fig. \ref{fig-4}.}
\label{fig-5}  
\end{figure}

The results of numerical analyses are shown in Figs. \ref{fig-2}-\ref{fig-5}.
Fig. \ref{fig-2} (a) and (b) show three curves corresponding to 
$\Delta_{B_3}=5,10,15$ for $B_3=500$ and $300$ GeV in the case (i), 
respectively. 
The darkest (darker) solid lines correspond $\Delta_{B_3}=5$ ($10$).
The dotted (red) curve is $m_h=114.4.$ GeV 
and the shaded (yellow) region corresponds to the region 
with $m_h\geq114.4$ GeV. 
In these figures, we fix $b_1=a_1=1$ and 
$b_2\simeq 4.19$ for (a) and $b_2\simeq 4.01$ for (b). 
These values of $b_2$ lead to 
$\Delta_{B_3}=10$ when $b_1=a_1=a_2=a_3=1$.\footnote{There is another value of 
$b_2$, which is negative and its absolute value is similar for positive one, to 
lead to $\Delta_{B_3}=10$ when $b_1=a_1=a_2=a_3$=1. In this work, we focus on 
only a positive value of solution but our results are not modified for a 
negative one.} 
These figures mean how mush stable the region with $\Delta_{B_3}=10$
is in the $(a_2,a_3)$ plane, when $b_2$ is fixed such that 
$\Delta_{B_3}=10$ is realized for $b_1=a_1=a_2=a_3=1$.
We find from
 Fig. \ref{fig-2} (a) that $a_2\lseq5$ and $a_3\lseq2$ are required to realize 
$\Delta_{B_3}\sim10$. 
Fig. \ref{fig-2} (b) shows that 
these upper bounds of both $a_2$ and $a_3$ are raised for $B_3=300$ GeV. 
It is seen from \eqref{fine-GUT} 
that the widths among three lines become wider  
as the $B_3$ becomes lower. The 
lower bound for $a_2$ and $a_3$ are evaluated as 
$(a_2,a_3)\gseq(-45,-10)$ for $B_3=500$ GeV and $(-40,-9)$ for $300$ GeV. These
 results are insensitive to a value of $a_1$, even if 
$a_1$ is larger  such as 
$a_1\sim\mathcal{O}(10)$. 
Figure \ref{fig-2} (c) and (d) correspond to the case 
of $(b_2,B_3)=(1,500\mbox{ GeV})$ and $(b_2,B_3)=(1,300\mbox{ GeV})$, 
respectively. The lower bounds for $a_2$ are raised to  $a_2\gseq-10$. It 
can be also found that the favorable region is $a_3\gseq-15$. 

{}Figures \ref{fig-3} and \ref{fig-4} show the results of the same analyses 
as case (i) for the cases (ii) and (iii), respectively, but (a) and (c) for 
$B_3=1$ TeV and (b) and (d) for $B_3=500$ GeV in Fig. \ref{fig-4}. For the
 messenger scale of $M=10^6$ GeV, there is no region corresponding to  
$\Delta_{B_3} \sim 10$ and $m_h\geq114.4$ GeV when the gluino 
mass is relatively light such as $B_3\sim300$ GeV. 
Fig. \ref{fig-5} 
corresponds to the enlargement of Fig. \ref{fig-4}. 
All favorable regions shown in figures also satisfy 
the experimental bound of the top squark mass, 
$m_{\tilde{t}_1}\geq95.7$ GeV. The allowed regions 
become generally narrow as the messenger scale becomes lower. Especially, the 
values of $a_2$ and $a_3$ are constrained to only negative values for 
$(M,B_3,b_2)=(10^6\mbox{ GeV}, 500\mbox{ GeV}, 1)$ shown in Fig. \ref{fig-4} 
(d). This means that tachyonic scalar masses are required at the messenger scale
 to reduce the fine-tuning in the context of the GGM.

Toward for future model building of the GGM to relax the fine-tuning 
problem, we present a summary of a typical parameter space in Tables 
\ref{tab:parameter-GUT},\ref{tab:parameter-10-10},\ref{tab:parameter-10-6}.
When we fix as $a_3=1$, which is always allowed in all cases of the 
messenger scale, the favorable regions are obtained as 
 
\noindent (i) $M=\Lambda_{\mbox{{\scriptsize GUT}}}$
\begin{eqnarray}
  &&0\lseq b_2\lseq7\mbox{ for }-100\lseq a_2\lseq 40\mbox{ and }B_3=500\mbox{ GeV}, \\
  &&0\lseq b_2\lseq6.5\mbox{ for }-100\lseq a_2\lseq 40\mbox{ and }B_3=300\mbox{ GeV}, 
 \end{eqnarray}

\noindent (ii) $M=10^{10}$ GeV
 \begin{eqnarray}
  && 2\lseq b_2\lseq8\mbox{ for }a_3=1,~-100\lseq a_2\lseq20,\mbox{ and }
     B_3=500\mbox{ GeV}, \\
  && 0\lseq b_2\lseq6.5\mbox{ for }a_3=1,~-100\lseq a_2\lseq-10, \mbox{ and }
     B_3=300\mbox{ GeV},
 \end{eqnarray}

\noindent (iii) $M=10^6$ GeV
 \begin{eqnarray}
  && 5\lseq b_2\lseq11\mbox{ for }a_3=1,~-100\lseq a_2\lseq5,\mbox{ and }
     B_3=1000\mbox{ GeV}, \\
  && 5\lseq b_2\lseq10\mbox{ for }a_3=-1,~-100\lseq a_2\lseq-10, \mbox{ and }
     B_3=500\mbox{ GeV}.
 \end{eqnarray}

Our results show that a certain ratio between the gluino mass 
and wino mass is favorable.
Also, the tachyonic initial condition for stop masses at 
the messenger scale would be favorable, in particular in 
the low messenger scale scenario.
For $M< 10^6$ GeV, the favorable region corresponds to 
only negative values of both $a_2$ and $a_3$.
The $A$-term $A_t$ plays a role in this result.
Its initial value vanishes at $M$, i.e. $A_t(M)=0$, 
and its value at $M_Z$ is generated by RG effect as 
Eqs.~\eqref{GUT-A},\eqref{10-A},\eqref{6-A}, which are 
determined mainly by $B_3$ and $B_2$.
However, a value of $|A_t(M_Z)|$ at $M_Z$ is smaller as 
the messenger scale becomes lower, because the RG effects become 
smaller.
On the other hand, a large value of the stop mixing 
$|A_t/m_{\tilde t}|$ is favorable to increase the Higgs mass, $m_h$.
Thus, if a value of $|A_t(M_Z)|$ is small, 
we have to decrease a value $m_{\tilde t}$ to obtain a large 
stop mixing $|A_t/m_{\tilde t}|$.
That can be realized by imposing the tachyonic 
initial condition of the stop mass at $M$.

%\noindent (i) $M=\Lambda_{\mbox{{\scriptsize GUT}}}$
\begin{table}[h]
\begin{center}
\begin{tabular}{c|c|c|c|c}
\hline
$B_3$ [GeV] & $b_2$ & $a_2$ & $a_3$ & Figure\\
\hline
\hline
$500$ & $4.19$ & $-45\lseq a_2\lseq5$ & $-10\lseq a_3\lseq2$ & 2 (a)\\
\hline
$300$ & $4.01$ & $-40\lseq a_2\lseq10$ & $-9\lseq a_3\lseq5$ & 2 (b)\\
\hline
$500$ & $1$ & $-10\lseq a_2\lseq50$ & $-15\lseq a_3\lseq0$ & 2 (c) \\
\hline
$300$ & $1$ & $-10\lseq a_2\lseq50$ & $-12\lseq a_3\lseq3$ & 2 (d) \\
\hline
\end{tabular}
\end{center}
\caption{Favorable parameter regions for (i) 
$M=\Lambda_{\mbox{{\scriptsize GUT}}}$.}
\label{tab:parameter-GUT}
\end{table}

%\noindent (ii) $M=10^{10}$ GeV
\begin{table}[h]
\begin{center}
\begin{tabular}{c|c|c|c|c}
\hline
$B_3$ [GeV] & $b_2$ & $a_2$ & $a_3$ & Figure\\
\hline
\hline
$500$ & $4.12$ & $-35\lseq a_2\lseq15$ & $-5\lseq a_3\lseq0$ & 3 (a)\\
\hline
$300$ & $3.86$ & $-35\lseq a_2\lseq-10$ & $-5\lseq a_3\lseq-1$ & 3 (b)\\
\hline
$500$ & $1$ & $-10\lseq a_2\lseq10$ & $-6\lseq a_3\lseq-3$ & 3 (c) \\
\hline
$300$ & $1$ & $-12\lseq a_2\lseq-2$ & $-5\lseq a_3\lseq-4$ & 3 (d) \\
\hline
\end{tabular}
\end{center}
\caption{Favorable parameter regions for (ii) 
$M=10^{10}$ GeV.}
\label{tab:parameter-10-10}
\end{table}

%\noindent (iii) $M=10^6$ GeV
\begin{table}[h]
\begin{center}
\begin{tabular}{c|c|c|c|c}
\hline
$B_3$ [GeV] & $b_2$ & $a_2$ & $a_3$ & Figure\\
\hline
\hline
$1000$ & $4.10$ & $-20\lseq a_2\lseq-5$ & $-2\lseq a_3\lseq0$ & 4 (a)\\
\hline
$500$ & $3.93$ & $-20\lseq a_2\lseq-10$ & $-2\lseq a_3\lseq-1$ & 4 (b)\\
\hline
$1000$ & $1$ & $-5\lseq a_2\lseq5$ & $-2\lseq a_3\lseq-1$ & 4 (c) \\
\hline
$500$ & $1$ & $-5\lseq a_2\lseq-3$ & $-3\lseq a_3\lseq-2$ & 4 (d) \\
\hline
\end{tabular}
\end{center}
\caption{Favorable parameter regions for (ii) 
$M=10^{6}$ GeV.}
\label{tab:parameter-10-6}
\end{table}

We also give the mass spectra of gluino, wino, and stop for typical 
parameters of the favorable regions in Table \ref{tab-1}.
\begin{table}[t]
\begin{center}
\begin{tabular}{|c|c|c|c|c|c|c|c|}
\hline
$M$ [GeV]& $B_3$ [GeV] & $a_3$ & $a_2$ & $b_2$ & $M_3$ [GeV] & $M_2$ [GeV] &
$m_{\tilde{t}}$ [GeV] \\
\hline
\hline
$2\times10^{16}$ & $500$ & $1$ & $-50$ & $5.71$ & $1500$ & $2450$ & $1180$ \\
\hline
$2\times10^{16}$ & $500$ & $-1$ & $-50$ & $4.22$ & $1500$ & $1810$ & $863$ \\
\hline
$2\times10^{16}$ & $500$ & $-1$ & $-1$ & $1.31$ & $1500$ & $563$ & $865$ \\
\hline
$2\times10^{16}$ & $500$ & $-1$ & $40$ & $1.42$ & $1500$ & $609$ & $803$ \\
\hline
$2\times10^{16}$ & $300$ & $1$ & $-50$ & $5.59$ & $900$ & $1440$ & $722$ \\
\hline
$2\times10^{16}$ & $300$ & $1$ & $1$ & $4.01$ & $900$ & $1030$ & $616$ \\
\hline
$2\times10^{16}$ & $300$ & $1$ & $30$ & $2.63$ & $900$ & $677$ & $522$ \\
\hline
$2\times10^{16}$ & $300$ & $-1$ & $-50$ & $5.36$ & $900$ & $1380$ & $704$ \\
\hline
$2\times10^{16}$ & $300$ & $-1$ & $1$ & $3.67$ & $900$ & $947$ & $615$ \\
\hline
$2\times10^{16}$ & $300$ & $-1$ & $30$ & $2.01$ & $900$ & $517$ & $555$ \\
\hline
$10^{10}$ & $500$ & $1$ & $-50$ & $6.48$ & $1500$ & $2790$ & $1280$ \\
\hline
$10^{10}$ & $500$ & $1$ & $1$ & $4.12$ & $1500$ & $1770$ & $1230$ \\
\hline
$10^{10}$ & $500$ & $1$ & $10$ & $3.58$ & $1500$ & $1520$ & $1210$ \\
\hline
$10^{10}$ & $500$ & $-1$ & $-50$ & $6.04$ & $1500$ & $2590$ & $1150$ \\
\hline
$10^{10}$ & $500$ & $-1$ & $1$ & $3.34$ & $1500$ & $1440$ & $1110$ \\
\hline
$10^{10}$ & $500$ & $-1$ & $10$ & $2.55$ & $1500$ & $1100$ & $1100$ \\
\hline
$10^{10}$ & $300$ & $1$ & $-30$ & $5.50$ & $900$ & $1420$ & $755$ \\
\hline
$10^{10}$ & $300$ & $-1$ & $-50$ & $5.89$ & $900$ & $1510$ & $682$ \\
\hline
$10^{10}$ & $300$ & $-1$ & $-10$ & $3.83$ & $900$ & $987$ & $669$ \\
\hline
$10^6$ & $1000$ & $1$ & $-50$ & $8.32$ & $3000$ & $7150$ & $2420$ \\
\hline
$10^6$ & $1000$ & $-1$ & $-50$ & $7.59$ & $3000$ & $6520$ & $1800$ \\
\hline
$10^6$ & $1000$ & $-1$ & $1$ & $2.21$ & $3000$ & $1900$ & $1800$ \\
\hline
$10^6$ & $500$ & $-1$ & $-30$ & $5.97$ & $1500$ & $2570$ & $897$ \\
\hline
\end{tabular}
\end{center}
\caption{Mass spectra of gluino, wino, and stop in typical parameter space.}
\label{tab-1}
\end{table}
We find that the smallest masses of wino and stop are realized in the case (i) 
with $B_3=300$ GeV, ${a}_3=-1$, and ${a}_2=30$ as $M_2\simeq517$ 
GeV and $m_{\tilde{t}}\simeq 555$ GeV. On the other hand, the largest masses of 
wino and stop are given in the case (iii) with $B_3=10^3$ GeV, ${a}_3=1$ 
and ${a}_2=-50$ as $M_2\simeq7150$ GeV and $m_{\tilde{t}}\simeq2420$ GeV.

\section{$\mu-B$ problem}

Here, we comment on the $\mu$-term and $B$-term.
How to generate the $\mu$-term and $B$-term is another important issue.
Within the framework of the gauge mediation, 
a simple mechanism to generate the $\mu$-term would lead to 
\begin{eqnarray}\label{eq:B/mu}
\frac{\mu B}{\mu^2} = {\cal O}(16\pi^2).
\end{eqnarray}
This ratio would cause a problem if 
\begin{eqnarray}\label{eq:mu-hd}
\mu^2 \sim m^2_{H_u}(M_Z), ~m^2_{H_d}(M_Z).
\end{eqnarray}
When both \eqref{eq:B/mu} and \eqref{eq:mu-hd} hold, 
 we could not realize the successful EW symmetry breaking.
That is often called the  $\mu-B$ problem of the 
gauge mediation.

However, in the previous section, 
we have studied models with spectra different from
Eq.~(\ref{eq:mu-hd}).
{}From the viewpoint of fine-tuning between $\mu^2$ and
$m^2_{H_u}(M_Z)$, 
the favorable spectrum is that 
$\mu, |m_{H_u}(M_Z)| = {\cal O}(100)$GeV and 
other SUSY breaking masses are of  order of a few TeV.
Indeed, if we can obtain the following hierarchy,
\begin{eqnarray}\label{eq:mu-B-mu-Hd}
  \mu^2\sim m_{H_u}^2\ll \mu B\ll m_{H_d}^2,
 \end{eqnarray}
we can realize the successful EW symmetry breaking.
It has been already pointed out in \cite{Csaki:2008sr} 
that the above hierarchy would be favorable in 
the gauge mediation.
Also, such a pattern has been studied within the framework of 
the TeV scale mirage scenario \cite{Choi:2006xb}, i.e. the 
mass pattern II.

This pattern of hierarchy can be 
realized in our analyses. A relatively large 
$B_2$ is favorable to obtain a large $m_{H_d}$ seen as in \eqref{GUT-mhd}, 
\eqref{10-mhd}, and \eqref{6-mhd}. For example, if we take $M=10^6$ GeV, 
$B_3=1$ TeV, $a_3=1$, $a_2=-50$, and $b_2\simeq8.32$, which lead to 
$\Delta_{B_3}=10$, $M_3\simeq3$ TeV, and $M_2\simeq7.15$ TeV, and 
$m_{\tilde{t}}=2.42$ TeV, we obtain
 \begin{eqnarray}\label{eq:mhd}
  m_{H_d}^2(M_Z)\simeq2.89^2
  \mbox{ TeV}^2.
 \end{eqnarray}
By using 
 \begin{eqnarray}
 \sin2\beta=\frac{2\mu B}{2|\mu|^2+m_{H_u}^2+m_{H_d}^2},
 \end{eqnarray}
with $\tan \beta =10$,
the above value of $m_{H_d}^2(M_Z) \simeq 2.89^2$TeV$^2$ 
determines the value of $\mu B$ as 
 \begin{eqnarray}
  \mu B \simeq911^2
  \mbox{ GeV}^2.
 \end{eqnarray}
That is, we have $\mu B/\mu^2 ={\cal O}(100)$ for $\mu \sim 100$ GeV.
Such a ratio $\mu B/\mu^2 $ could be realized by a simple mechanism to 
generate the $\mu$-term and $B$-term (\ref{eq:B/mu}).\footnote{
In this example, we use the large ratio of $|a_2/a_3|$, 
i.e. $a_3=1$ and $a_2=-50$, 
but realization of such a ratio may not be straightforward in
explict model building.
As another example, we take $M=10^{10}$ GeV, 
$B_3=500$ GeV, $a_3=1$, $a_2=1$, and $b_2\simeq 4.12$.
This example leads to $m_{H_d}(M_Z)\simeq 1.06$ TeV 
and $\mu B/\mu^2 ={\cal O}(10)$.
That would lead to the above hierarchy (\ref{eq:mu-B-mu-Hd}) 
although the gap of hierarchy would be smaller than the first example.}
Therefore, this parameter set, which relaxes the fine-tuning problem, 
would also be favorable from the viewpoint of 
the $\mu-B$ problem.

%%%%%%%%%%%%%%%%%
\section{Summary}
%%%%%%%%%%%%%%%%%
We have studied the fine-tuning problem in the context of general gauge 
mediation. Numerical analyses toward for relaxing the fine-tuning in the problem
 have been presented. We analysed the problem in typical three cases of the 
messenger scale, that is, GUT ($2\times10^{16}$ GeV), intermediate 
($10^{10}$ GeV), and relatively low energy ($10^6$ GeV) scales. In each case, 
the parameter space with less fine-tuning such as $10\%$ has been found. 
It has also been shown that the favorable region becomes narrow as the messenger 
scale becomes lower, especially, $-10\lseq a_2\lseq50$ and $-15\lseq a_3\lseq0$ 
are allowed for $B_3=500$ GeV and $b_1=b_2=a_1=1$ in the case (i), 
$-10\lseq a_2\lseq 10$ and $-6\lseq a_3\lseq -3$ for $B_3=500$ GeV and 
$b_1=b_2=a_1=1$ in the case (ii), and $-5\lseq a_2\lseq-3$ and 
$-3\lseq a_3\lseq -2$ for $B_3=500$ GeV and $b_1=b_2=a_1=1$ in the case (iii).
Our results imply that certain ratios between the gluino and wino
masses as well as scalar masses are favorable to relax 
the fine-tuning problem.
Also, 
tachyonic initial conditions of scalar masses 
are favored, in particular in the relatively low messenger scale
scenario.  
Furthermore, the type of spectra with $\mu \approx 100$ GeV and 
a few TeV of other SUSY breaking masses is also favorable from 
the viewpoint of the $\mu-B$ problem.
Thus, it would be important to construct explicit models, 
which realize certain ratios among gaugino and scalar masses.

\vskip 1cm
{\bf Note to be added}

While this paper was being completed, Ref.~\cite{Abel:2009ve}
appeared, where also fine tuning in the GGM was 
studied.

\subsection*{Acknowledgement}

T.~K. is supported in part by the Grant-in-Aid for 
Scientific Research No.~20540266 from the 
Ministry of Education, Culture, Sports, Science and Technology of Japan.
T.~K. and R.~T. are also supported in part 
by the Grant-in-Aid for the Global COE 
Program "The Next Generation of Physics, Spun from Universality and 
Emergence" from the Ministry of Education, Culture,Sports, Science and 
Technology of Japan.

%%%%%%%%%%%%%%%%%%%%%%%%%%%

\end{document}